\documentclass[twocolumn]{revtex4}
\usepackage{amsmath}
\usepackage{latexsym}
\usepackage{graphicx}
\usepackage{pifont}
\usepackage{color}
\usepackage[usenames,dvipsnames]{xcolor}

\newcommand{\req}[1]{Eq.\,(\ref{#1})}
\newcommand{\beqn}{\begin{equation}}
\newcommand{\eeqn}{\end{equation}}

\newcommand{\kur}{\kappa_4}

\begin{document}

\title{Baryon susceptibilities, nongaussian moments and the QCD critical
point}
\author{Jiunn-Wei Chen$^{1,2}$, Jian Deng$^{3}$ and Lance Labun$^2$}
\affiliation{$^{1}$ Department of Physics and National Center for Theoretical Sciences,
National Taiwan University, Taipei, 10617 Taiwan\\
$^2$ Leung Center for Cosmology and Particle Astrophysics (LeCosPA)\\
National Taiwan University, Taipei, 10617 Taiwan\\
$^3$ Key Laboratory of Particle Physics and Particle Irradiation (MOE),
School of Physics, Shandong University, Jinan 250100, China}
\date{18 October, 2014}

\begin{abstract}
We calculate model-independently the impact of the critical point on higher
order baryon susceptibilities $\chi_n$, showing how they depend on
fluctuations of the order parameter. Including all tree level diagrams, we
find new contributions to $\chi_4$ equally important to the kurtosis of the
order parameter fluctuations, and we characterize the kurtosis and other
nonguassian moments as functions on the phase diagram. Important features of this analysis are then confirmed by a Gross-Neveu model study with good agreement with other model studies as well as lattice and experimental data. This suggests the universality of the characteristic peak in baryon susceptibilities as a signal of the critical point.  We discuss leveraging measurements of different $\chi_n$ to extrapolate the location of the critical point.%
\end{abstract}

\maketitle



A major goal of QCD theory and heavy-ion collision (HIC) experiment is to
locate the critical end point in the chemical potential--temperature ($\mu\!-\!T$) plane\thinspace \cite{QCDphases}. It is the target of the beam
energy scan at RHIC and the future FAIR experiment, which are designed to
create and measure QCD matter at high temperature and density. Lattice
simulations are also developing methods to calculate properties of QCD
matter at $\mu \neq 0$\thinspace \cite{altlattice,Gavai:2010zn}, which
cannot be reached directly due to the sign problem.

The critical point itself is a second-order transition, characterized by
diverging correlation length $\xi $, due to vanishing mass of
the order parameter field $\sigma $. This fact, $m_{\sigma }^{-1}=\xi \rightarrow \infty $, is a statement about the two-point correlation
function of the $\sigma $ field, and we can use low energy effective
field theory to relate other correlation functions to the critical point and phase structure.  $\sigma $ correlations influence observables such as
baryon number fluctuations because the $\sigma $ couples like a mass term
for the baryons, meaning that the presence of $\sigma $ changes the baryon
energy\thinspace \cite{Stephanov:1998dy}. Thus our aim is to establish the
theory connection from the phase structure through $\sigma $ dynamics to
observables, here proton number fluctuations, which can be compared to
event-by-event fluctuations in HICs\thinspace \cite{Aggarwal:2010wy,Adamczyk:2013dal} and to lattice simulations\thinspace \cite{Gavai:2010zn}.

It is important to keep in mind that the QCD matter created in HICs is
dynamic. The measured data in general integrate properties from the initial
state and expansion dynamics, and they may not represent equilibrium
properties of QCD matter at the freeze-out $\mu ,T$, especially if the
fireball has passed near the critical point~\cite{slowing}. Assuming the
departure from equilibrium is small, we interpret the freeze-out data as
approximate measurements of the phase diagram, which can be compared with
theory and lattice predictions to help locate the critical point.

The fluctuation observables compared between HICs and lattice simulations
are ratios of baryon susceptibilities 
\begin{equation}
m_{1}=\frac{T\chi _{3}}{\chi _{2}},\quad m_{2}=\frac{T^{2}\chi _{4}}{\chi
_{2}},\quad \chi _{n}=\frac{\partial ^{n}\ln \mathcal{Z}}{\partial \mu ^{n}}
\label{chin}
\end{equation}%
with the volume dependence eliminated in the ratios.  More precisely, HICs measure proton fluctuations, which are shown to directly reflect the baryon fluctuations, because the order parameter field, the scalar $\sigma$, is an isospin singlet \cite{Hatta:2003wn}.
From here, one approach is model independent, considering the partition function as a path integral over $\sigma $, 
$\mathcal{Z}=\int \!\mathcal{D}\sigma \ e^{-\Omega \lbrack \sigma ]/T}$, 
and the effective potential of the Landau theory $\Omega[\sigma]$ contains the phase
structure in its coefficients. However those parameters are not determined
by the theory.  Previously this has been used to search for
dominant contributions to $\chi _{n}$ close to the critical point~\cite{Stephanov:2008qz,Stephanov:2011pb}. Another approach is to evaluate $\ln \mathcal{Z}$ in a QCD-like model, such as NJL~\cite{Asakawa:2009aj}, to gain
predictive power of $\chi _{n}$\ as functions on the phase diagram. We
pursue both approaches to put the model independent results into the context
of the global phase diagram.

We analyze a general polynomial form of the effective potential $\Omega[\sigma]$. We derive the $\chi _{n}$ as functions of the $\sigma $
fluctuation moments $\langle \delta \sigma ^{k}\rangle $, extracting new,
equally important contributions to $m_{2}$ in additional to the $\sigma $
field kurtosis $\kappa_{4}$, studied by \cite{Stephanov:2011pb}. We show
that negative $\kappa_{4}$ is restricted to the normal phase, and thus these new
contributions are necessary to understand recent HIC and lattice results for 
$m_{2}$. Our model independent results are corroborated with quantitative
study of the 1+1 dimensional Gross-Neveu (GN) model, revealing remarkably
good qualitative agreement with both other model studies~\cite{Asakawa:2009aj} as well as the experimental data. This consistency suggests that those features of our findings are model-independent.

We begin with the effective potential for the order parameter field, 
\begin{equation}
\Omega \lbrack \sigma ]=\int d^{3}x\left( -\!J\sigma +\frac{g_{2}}{2}\sigma
^{2}+\frac{g_{4}}{4}\sigma ^{4}+\frac{g_{6}}{6}\sigma ^{6}+\cdots \right)
\label{Veffglobal}
\end{equation}%
with coefficients $g_{2n}$ functions of temperature and chemical potential,
determining the phase diagram. Focusing on long range correlations, we
consider only the zero momentum $\vec{k}=0$ mode, and so do not write the
kinetic energy term $(\vec{\nabla}\sigma )^{2}$ here \cite{Stephanov:2008qz}%
. With the explicit symmetry breaking parameter $J\rightarrow 0$, the point
where $g_{2}=g_{4}=0$ is the tricritical point (TCP), separating the second
order transition line for $g_{4}>0$ from the first order line for $g_{4}<0$.
When $J\neq 0$, the second order line disappears into a crossover transition
through which the $\sigma $ minimum $\langle \sigma \rangle \equiv v$
changes smoothly as a function of temperature, and the TCP becomes a
critical end point (CEP).

Fluctuations of the order parameter field obey an effective potential
obtained by first minimizing the potential Eq.\thinspace (\ref{Veffglobal})
and then Taylor expanding around $v$, yielding 
\begin{equation}
\Omega \lbrack \delta \sigma ]-\Omega _{0}=\int d^{3}x\left( \frac{m_{\sigma
}^{2}}{2}\delta \sigma ^{2}+\frac{\lambda _{3}}{3}\delta \sigma ^{3}+\frac{%
\lambda _{4}}{4}\delta \sigma ^{4}+\cdots \right)  \label{Veffflucns}
\end{equation}%
with $\delta \sigma (x)=\sigma (x)-v$. The constant 
$\Omega _{0}\equiv\Omega[\sigma \!=\!v]$ 
does not influence the fluctuations, but does appear in the observables
corresponding to the mean field contribution.  The vev $v$ satisfies the gap equation $v(g_{2}+g_{4}v^{2}+g_{6}v^{4})=J$, and depends on $\mu ,T$ through the $g_{2n}$.

Calculating $\mu $-derivatives of the partition function gives an explicit
relation between susceptibilities $\chi _{n}$ and $\delta \sigma $
fluctuations. Starting with the second order, 
\begin{equation}
T^{2}\chi _{2}=T^{2}\frac{\partial ^{2}\ln \mathcal{Z}}{\partial \mu ^{2}}%
=-T\langle \Omega ^{\prime \prime }\rangle +\langle (\Omega ^{\prime
})^{2}\rangle -\langle \Omega ^{\prime }\rangle ^{2}  \label{dPdmu2}
\end{equation}%
where $\langle f\rangle =\mathcal{Z}^{-1}\int \mathcal{D}\sigma \ f\ e^{-\Omega /T}$ is the expectation value of the function $f$ including $\sigma $ fluctuations. The prime indicates differentiation with respect to $\mu $, 
\begin{equation}\label{aijexp}
\frac{\partial ^{k}\Omega }{\partial \mu ^{k}}=\int d^{3}x\left(
a_{k0}+a_{k1}\delta \sigma +a_{k2}\delta \sigma ^{2}+\cdots \right) .
\end{equation}%
The first term $a_{k0}$ is the mean-field contribution from differentiating $\Omega_0$.  The linear term arises from the $\mu$-dependence of the vev $v$. 

 Plugging these derivatives into Eq.\thinspace (\ref{dPdmu2}), we keep all \textquotedblleft tree-level\textquotedblright\ contributions, where the power of the correlator is less than or equal to order of the $\mu$-derivative. This means that the expectation value of a product of correlators at different points is equal to the product of expectation values of correlators formed by making all possible contractions of $\delta\sigma $ at different points. The combination 
$\langle (\Omega ^{\prime})^{2}\rangle -\langle \Omega ^{\prime }\rangle ^{2}$ 
cancels disconnected  diagrams. Applying these rules, 
\begin{equation}
T^{2}\chi _{2}=-VTa_{20}+V^2a_{11}^{2}\langle\delta\sigma^{2}\rangle
\label{dPdmu2flucns}
\end{equation}%
A diagrammatic method helps to organize these calculations and distinguish
loops arising from contractions.   So far \req{dPdmu2flucns} is just the usual second moment of particle number, here expanded in terms of the fluctuations of the $\delta\sigma$ field.

Applying this procedure, the higher order susceptibilities are 
\begin{align}  \label{dPdmu3sigma} 
T^{3}\chi _{3}=& 
-VT^{2}a_{30}+3V^2Ta_{11}a_{21}\langle \delta \sigma^{2}\rangle 
\\ \notag  & 
-V^3a_{11}^{3}\langle \delta \sigma^{3}\rangle -6V^{3}a_{11}^{2}a_{12}\langle \delta \sigma^{2}\rangle ^{2}  
\end{align}
and 
\begin{align}   \label{dPdmu4sigma} 
T^{4}\chi _{4}=& -VT^{3}a_{40}+V^2T^{2}(4a_{31}a_{11}+3a_{21}^{2})\langle \delta\sigma^{2}\rangle  \\
& -6V^3Ta_{21}a_{11}^{2}\langle\delta\sigma^{3}\rangle 
+V^4a_{11}^{4}\big(\langle \delta \sigma^{4}\rangle -3\langle \delta \sigma^{2}\rangle ^{2}\big)  \notag \\
& -12V^{3}T(a_{22}a_{11}^{2}+2a_{21}a_{11}a_{12})\langle \delta \sigma^{2}\rangle ^{2}  \notag \\
& +24V^{4}(2a_{11}^{2}a_{12}^{2}+a_{11}^{3}a_{13})\langle \delta \sigma^{2}\rangle ^{3}  \notag \\
& +24V^{4}a_{11}^{3}a_{12}\langle \delta \sigma^{3}\rangle \langle\delta \sigma^{2}\rangle   \notag
\end{align}
Each factor of $V$ comes from the $d^3x$ integration in \req{aijexp}, and after inserting the expressions for $\langle\delta\sigma^{k}\rangle $, each $\chi_2,\chi_3,\chi_4\propto V$.  The fluctuation moments $\langle\delta\sigma^{k}\rangle $ are derived by
functional differention of Eq.\thinspace (\ref{Veffflucns}), 
\begin{align}
\kappa _{2}& =\langle \delta \sigma ^{2}\rangle =\frac{T}{V}\xi ^{2},\quad
\kappa _{3}=\langle \delta \sigma ^{3}\rangle =-2\lambda _{3}\frac{T^{2}}{V^{2}}\xi ^{6}  \label{kappa4} \\
\kappa _{4}& =\langle \delta \sigma ^{4}\rangle -3\langle \delta \sigma
^{2}\rangle ^{2}=6\frac{T^{3}}{V^{3}}\big(2(\lambda _{3}\xi )^{2}-\lambda
_{4}\big)\xi ^{8}
\end{align}
The point is that the $a_{jk}$ coefficients weight how the $\delta\sigma$ correlations contribute to the higher order susceptibilities, $\chi_2,\chi_3...$.  Moveover, the $a_{jk}$ have their own $\xi$ dependence, which can be estimated analytically and model-independently, as well as compared with model studies.   For example, we find that $a_{11}=m^2\partial v/\partial\mu$ scales $\sim\xi^{-1}$  near the critical point.  To compare to a given solvable model (such as the GN model below), the coupling constants $m_\sigma^2,\lambda_3,\lambda_4...$ are calculated from the model's effective potential and then their $\mu$-derivatives evaluated yielding $a_{kj}$ coefficients.   

The third moment $\chi_{3}$ has been studied in the NJL model and found to be negative around the phase boundary \cite{Asakawa:2009aj}. In agreement with power-counting $\xi $ only in the $\delta\sigma$ correlators~\cite{Stephanov:2008qz}, the behavior of $m_{1}$ near the critical point can be  explained by focusing on $\langle\delta\sigma^{3}\rangle $ and hence the function $\kappa_{3}(\mu ,T)$: In this case, estimating the $\xi$ dependence of the $a_{jk}$ coefficients in \req{dPdmu3sigma} reveals that the $a_{11}^3\kappa_3$ term scales with the largest positive power of $\xi$.

\begin{figure}[t]
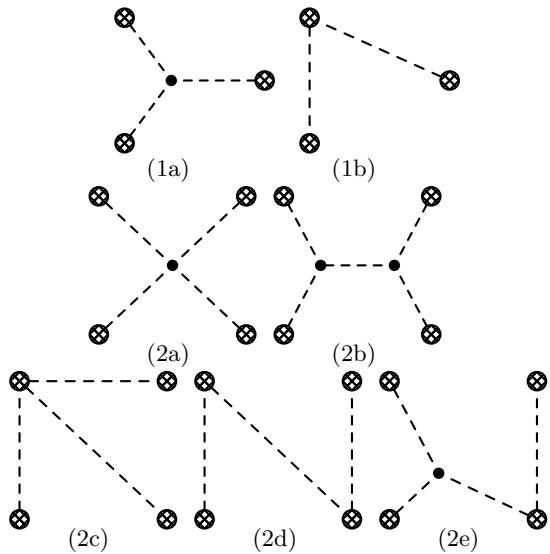

\begin{picture}(200,210)
\put(30,150){\includegraphics[width=70pt]{chi3lambda3.1}}
\put(55,140){(1a)}
\put(100,150){\includegraphics[width=70pt]{chi3xi2.1}}
\put(125,140){(1b)}
\put(30,80){\includegraphics[width=70pt]{chi4lambda4.1}}
\put(55,70){(2a)}
\put(100,80){\includegraphics[width=70pt]{chi4lambda3sq.1}}
\put(125,70){(2b)}
\put(0,10){\includegraphics[width=70pt]{chi4xi31.1}}
\put(25,0){(2c)}
\put(70,10){\includegraphics[width=70pt]{chi4xi32.1}}
\put(95,0){(2d)}
\put(140,10){\includegraphics[width=70pt]{chi4lambda3xi.1}}
\put(165,0){(2e)}
\end{picture}
\caption{Diagrams 1a,1b give leading contributions to $\chi_3$.  Diagrams 2(a-e) are some of the leading contributions to $\chi_4$.  Omitted are diagrams involving multiple $\mu$ derivaties at the same point. }
\label{fig:diags}
\end{figure}

However, for $\chi_{4}$ there are many terms of the same (tree-level) order in the perturbation theory.  Taking into account the $\xi$ dependence of the coefficients, several contributions, including those represented by the diagrams in Fig. \ref{fig:diags} scale with the same power of $\xi$ as the $\kappa_{4}$ term.  Although fewer $\sigma$ propagators are visible in some of these diagrams, the coefficient functions $a_{11},a_{12},$ and $a_{13}$ all have important $m_{\sigma}$ dependence.  Looking at $\xi$-scaling, we find all three terms $\langle\delta\sigma^2\rangle^2$, $\langle\delta\sigma^2\rangle^3$, and $\langle\delta\sigma^2\rangle\langle\delta\sigma^3\rangle$ are approximately equally relevant as the $\kappa_4$ term.  These analyses are supported by separately evaluating these terms in the GN model.

\begin{figure}[t]
\includegraphics[width=0.45\textwidth]{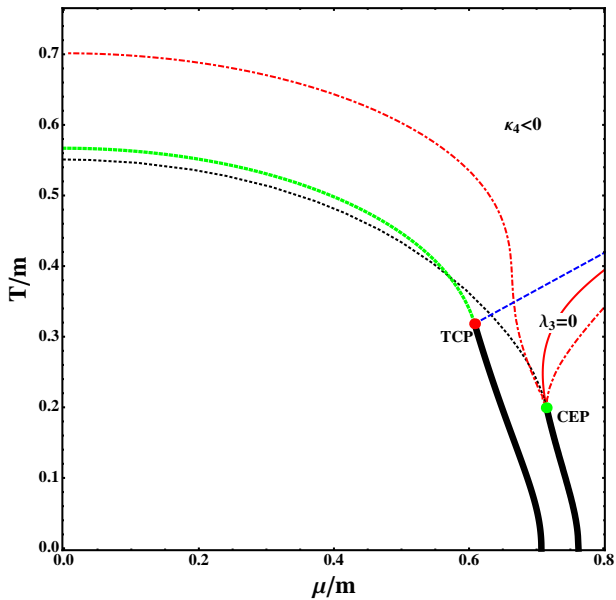}
\caption{Phase diagram of the GN model, with phase boundaries and TCP of the 
$\protect\gamma=0$ theory and the CEP of the $\protect\gamma=0.1$ theory.
The $\protect\kappa_4<0$ region of the $\protect\gamma=0$ is above the
second order (green) line and left of the dashed (blue) line that joins the boundary at the
TCP. The $\protect\kappa_4<0$ region of the $\protect\gamma=0.1$ theory is
delineated by the dot-dashed (red) line, and $\protect\lambda_3=0$ the solid
(red) line inside this region. }
\label{fig:GNphases}
\end{figure}

Next to see how the $\sigma $ fluctuations are impacted by the CEP, we
investigate $\kappa _{3}(\mu ,T)$ and $\kappa _{4}(\mu ,T)$ as functions on
the phase diagram. With $J\rightarrow 0$, the unbroken phase is where $%
\langle \sigma \rangle =0$, and in this case $\lambda _{2n}=g_{2n}$. Odd
terms are zero, in particular $\lambda _{3}\equiv 0$ in the unbroken phase,
and negative $\kappa _{4}$\ exists when $\lambda _{4}=g_{4}>0$\ above the
second order phase transition line. In the symmetry broken phase, 
\begin{align}
2(\lambda _{3}\xi )^{2}-\lambda _{4}& =\frac{4}{\sqrt{D}}\left( (g_{4}-2%
\sqrt{D})^{2}+D\right) ,  \label{kappa4broken} \\
& D=g_{4}^{2}-4g_{2}g_{6}>0\quad (J=0)  \notag
\end{align}%
$D$ is the algebraic discriminant obtained when solving the gap question for
the extrema, and it is positive in the broken phase, corresponding to real,
nontrivial ($\sigma \neq 0$) solutions. Therefore, with $J=0$, $\kappa_{4}$
is positive definite in the broken phase and the $\kappa_4<0$ region is
defined by the conditions $g_{2}>0$ and $g_{4}>0$ occuring only in the
unbroken phase. For concreteness, this is illustrated in the GN model,
Figure \ref{fig:GNphases}.

Turning on $J\neq 0$ produces a continuous change in the $\lambda_i$. In
particular, the $\kappa_4=0$ lines, bounding the $\kappa_4<0$ region,
move continuously away from their $J=0$ limits, and continue to obey the
constraint ``remembered'' from $J=0$ theory.

To see this, first recall that the tricritical point anchors one corner of
the $\kappa_{4}<0$ region, and in the $J\neq 0$ theory, the critical end
point continues to do so\:\cite{Stephanov:2011pb}. The reason is that $\kappa_{4}$ Eq.\thinspace (\ref{kappa4}) has a local minimum where $\lambda_{3}=0$. For $J=0$, $\lambda_{3}=0$ holds throughout the unbroken phase, but for
any fixed $J\neq 0$, the relation $\lambda_{3}(g_{2},g_{4},...)=0$ is an
equation whose solution defines a line in the $\mu -T$ plane. The 
$\lambda_{3}=0$\ line must pass through the critical end point. Differing
trajectories of the phase boundary and $\lambda_{3}=0$ line are seen in the
GN model, Figure \ref{fig:GNphases}.

The critical end point is located by the conditions $m_{\sigma}^{2}=\lambda_{3}=0$, which means the coefficients $g_{2n}$ satisfy~\cite{Stephanov:1998dy} 
\begin{equation}
g_{2}=5g_{6}v^{4},~~g_{4}=-\frac{10}{3}g_{6}v^{2},~~
v^{5}=\frac{3}{8}\frac{J}{g_{6}}\quad @\,\mathrm{CEP}  \label{CEPg2n}
\end{equation}%
The vev $v=\langle \sigma \rangle $ is nonzero, as expected, and as the
symmetry breaking is turned off $J\rightarrow 0$ these equations return to
their $J=0$ limits. Since $v^{2}>0$, the CEP always shifts to the
\textquotedblleft southeast\textquotedblright , into the fourth quadrant
relative to the TCP of the $J=0$ theory at $g_{2}=g_{4}=0$.

To locate the $\lambda_{3}=0$ line, relax condition on $m_{\sigma }^{2}$ to
find that $\lambda_{3}=0$ is the set of points satisfying 
\begin{equation}
g_{2}=\frac{7}{3}g_{6}v^{4}+\frac{J}{v},\quad g_{4}=-\frac{10}{3}g_{6}v^{2}
\label{lambda3zero}
\end{equation}%
The $\lambda_{3}=0$ line leaves the CEP parallel to the first order line,
and hence proceeds in the direction of decreasing $g_{4}$. With $g_{2}>0$
and $g_{4}<0$ near the critical point (Eq.\thinspace (\ref{CEPg2n})), the
relation Eq.\thinspace (\ref{lambda3zero}) requires that $v$ decreases along
the $\lambda_{3}=0$ line. In the high $T$ limit, $v\rightarrow 0$, so that
the $\lambda_{3}=0$ line asymptotes to $g_{4}=0$ from below. Thus, from
Eq.\thinspace (\ref{lambda3zero}) we deduce that $\lambda_{3}=0$ typically
cannot proceed close to the $\mu =0$ axis, since that would require that the
tricritical point of the $J=0$ theory is near the $\mu =0$ axis. The $\kappa_{4}<0$-region must migrate toward higher $T$ and $\mu $ with the critical
end point.

In the high $T$, low $\mu$ behaviour of $\kappa_4$ is given by expanding for
small $v$: 
$2(\lambda_3\xi)^2-\lambda_4= -g_4+2\big(\frac{(3g_4)^2}{g_2}-5g_6\big)v^2+
\mathcal{O}(v^4)$
which is valid where $g_2,g_4>0$, far away from the lines where $g_2,g_4$
vanish. Approaching from high $T$, $\kappa_4$ starts out negative just as in
the $J=0$ theory, and becomes positive just where the vev $v$ becomes large
enough that the second term starts to win over the first. Therefore, as the
magnitude of explicit breaking increases enhancing the order parameter, the $%
\kappa_4=0$ line and $\kappa_4<0$ region move farther from the phase
boundary.

We demonstrate the features derived above in the phase diagram and susceptibilities of the GN model. The fermion number susceptibilites behave very similarly to other models such as PNJL~\cite{Skokov:2011rq}. The GN model comprises $N$ fermions in 1 spatial dimension with bare mass $m_{0}$ and a four-fermion interaction $\propto g^{2}$, and in the large $N$ limit has a rich phase structure~\cite{Schnetz:2005ih}. The physical mass $m$ is given by $m\gamma =(\pi/Ng^{2})m_{0}$ where $\gamma =\pi /(Ng^{2})-\ln \Lambda /m$ is the parameter controlling the magnitude of explicit symmetry breaking. At small $\mu ,T$, there is a chiral condensate $\langle \bar{\psi}\psi \rangle $ and the order parameter is the effective mass 
$M=m_{0}-g^{2}N\langle \bar{\psi}\psi \rangle $. 
The effective potential is a function of $M$, and we focus on the region above and on the low $\mu $ side of the critical point~\cite{Schnetz:2005ih}

The phase diagram behaves as described model-independently: For $\gamma\rightarrow 0$, there is a tricritical point and second order line extending
to the $\mu =0$ axis. For $\gamma \neq 0$, the second order line vanishes
into a crossover and the critical end point shifts increasingly to the
\textquotedblleft southeast\textquotedblright\ away from the former
tricritical point. Figure~\ref{fig:GNphases} compares the phase diagrams of
the GN model for $\gamma =0$ and $\gamma =0.1$. For $\gamma \neq 0$ the
phase boundary is determined as the peak in the chiral susceptibility, 
\begin{equation}
\chi _{M}=\frac{\partial \langle \bar{\psi}\psi \rangle }{\partial m}=\frac{1%
}{m}\left( M-T\frac{\partial M}{\partial T}-\mu \frac{\partial M}{\partial
\mu }\right)  \label{chiM}
\end{equation}%
as is used in lattice QCD studies \cite{chiMlattice}. The phase boundary
stays near the critical line of the $\gamma =0$ theory, which is robust for
different values of $\gamma $. All our results are shown in units of $m=1$.

For $\gamma=0$, the $\kur<0$ region is delineated by the
second-order line and the $g_4=0$ line. For $\gamma=0.1$, it is delineated
by the dot-dashed line with a cusp at the CEP. Varying $\gamma$, we see that
the $\kur<0$  region evolves continuously as a function of $\gamma$
from its $\gamma\to 0$ limit. The $\lambda_3=0$ line leaves the CEP parallel
to the first order line, and the $\kappa_4<0$ region is approximately
symmetric around it very near the CEP. However, the $\lambda_3=0$ line then
asymptotes to the $g_4=0$ line, which pulls the $\kappa_4<0$ region away
from the phase boundary.

\begin{figure}[t]
\includegraphics[width=0.45\textwidth]{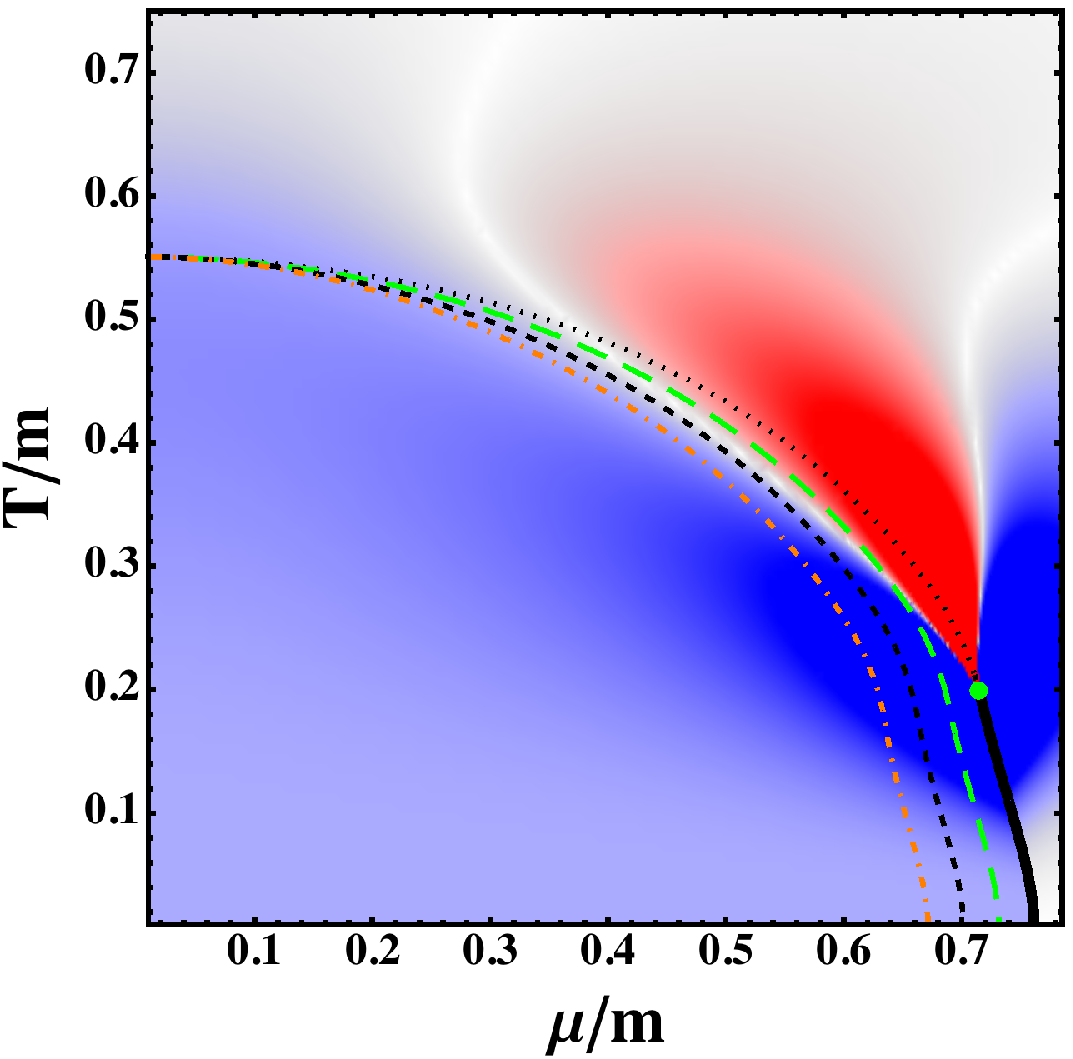}\newline
\includegraphics[width=0.45\textwidth]{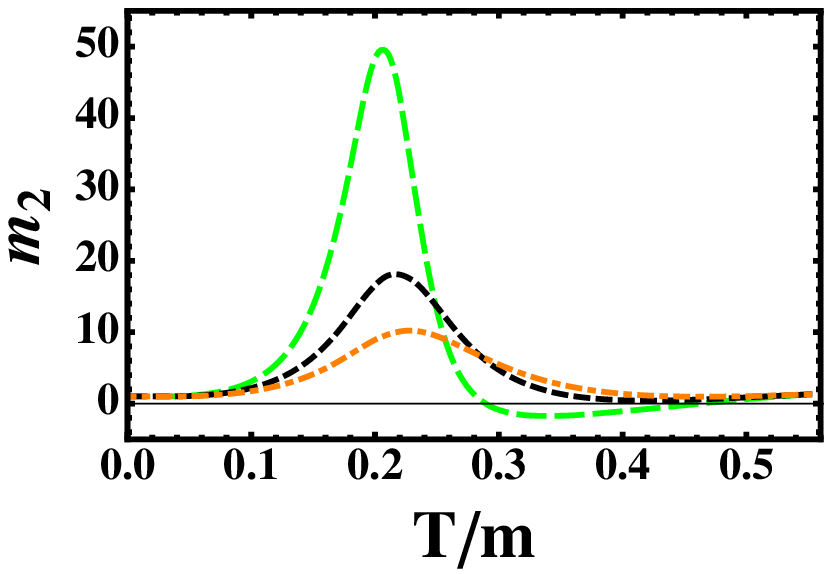}
\caption{Upper frame: Density plot of $m_2$ in the $\protect\mu-T$ plane
with $\protect\gamma=0.1$. The white lines indicate where $m_2=0$ and in the
(red) wedge between these lines $m_2<0$. The first order line is the solid
heavy line, and the crossover line is the dotted line, determined by the max
of Eq.\,(\protect\ref{chiM}). The dashed lines are hypothetical freeze-out
curves, color-coded to correspond to the lines in the lower frame. }
\label{fig:c2}
\end{figure}

We plot $m_2$ on the phase diagram in Figure \ref{fig:c2}. The negative $m_2$
region forms a wedge opening up from the CEP and extends deeper across the
phase boundary than the $\kappa_4<0$ region. Negative $m_2$ could be
accessible to freeze-out at $\mu<\mu_{\mathrm{CEP}}$, and the signature
would be a minimum followed by a rapid increase to a positive peak, as seen
in the (green) freeze-out curve closest to the phase boundary. Moving
freeze-out progressively away from the phase boundary, both the minimum and
maximum of $m_2$ decrease in magnitude. Thus it is possible $m_2$ is only
positive along the freeze-out curve (for example the lowest curve). Its
maximum provides a residual signal of proximity to the CEP, seeing that the
height of the peak decreases rapidly away from the phase boundary. Comparing
upper and lower frames of Fig. \ref{fig:c2}, we see that the peak in $m_2$
is always at a temperature higher (or $\mu$ lower) than the CEP.

Strikingly, the black line is in good qualitative agreement with lattice
and HIC results. However, non-monotonic behaviour of $m_{2}$ along a single
freeze-out line is insufficient to establish proximity to the CEP. Many
possible freeze-out curves can be drawn that cross several contours of
constant $m_{2}$ twice, and each will display a local maximum of $m_{2}$ as
a function of $\mu $ or the collision energy. For this reason, it will be
important to combine several probes of the phase diagram, and one way to
start is to compare $m_{2}$ and $m_{1}$

\begin{figure}[t]
\includegraphics[width=0.45\textwidth]{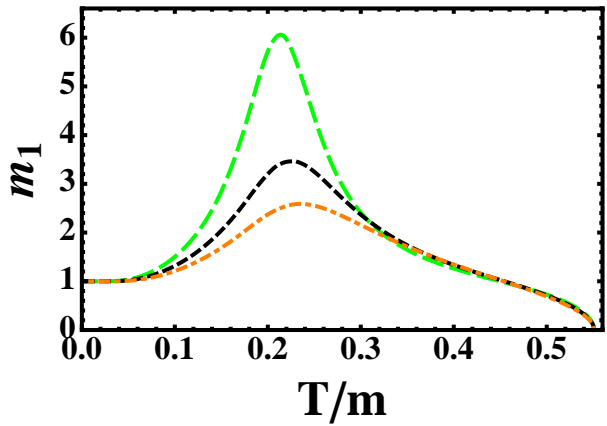}
\caption{$m_1$ along the hypothetical freeze-out lines given in the upper
frame of Fig.\protect\ref{fig:c2}. \\[-3mm]}
\label{fig:c1}
\end{figure}

In Figure \ref{fig:c1}, we plot $m_{1}$ along the same hypothetical
freeze-out curves. Like $m_{2}$ it displays a positive peak close the CEP,
and the magnitude of the maximum decreases for freeze-out lines farther away
from the phase boundary. Again, the peak is at higher temperature (lower $%
\mu $) than the CEP. This fact appears to be universal, as it is seen an
Ising-model evaluation of $\kappa _{3}$ and $\kappa _{4}$ similar to \cite{Stephanov:2011pb}. Despite many similarities, the topography of the peaks
in $m_{1}$ and $m_{2}$ differ in detail. Combining measurements of these two
observables along the freeze-out curve, we may be able to extract more
information about the CEP location.

To conclude, we have studied the fermion susceptibilities $\chi_2,\chi_3,\chi_4$ analytically using a low energy effective theory for the order parameter field and numerically using the Gross-Neveu model as an example system. The model-independent analysis shows that larger quark mass pushes the critical end point to higher $\mu$, and there are constraints on the position of the CEP relative to the tricritical point of the zero quark mass theory.

In agreement with previous work, nonmonotonic behaviour of $m_{1}$ and $m_{2} $ appears as a signal of the critical region in the phase diagram. 
{ Consistent with experimental data, we find $m_2$ first decreases as a function of chemical potential $\mu$, which is a remnant of the $m_2<0$ region above the critical point.  Seeing a large peak in $m_2$ at larger $\mu$/smaller $\sqrt{s}$ would support this explanation of the data.  However, it is necessary to accumulate as much corroborating evidence as possible to preclude false positive, and we note in this same region, $m_1$ is also expected to peak and decrease again.  The peaks in $m_{1},m_{2}$ are typically not the point of closest approach, and the temperature of the peaks are ordered 
$T_{\mathrm{max},m_{1}}>T_{\mathrm{max},m_{2}}>T_{\mathrm{CEP}}$, a fact which might be leveraged to indicate the location of the critical point}.

To the extent that the fireball is near thermodynamic equilibrium at
freeze-out, the model independent features we find can be compared to
experiment. It may be possible to refine the predictions by taking
into account expansion dynamics\thinspace \cite{slowing}. More information
may be extracted from the experimental data by combining measurements of $m_{1}$ and $m_{2}$ along the single available freeze-out curve (and possibly
other curves available from lattice).

\vskip0.2cm 
\textit{Acknowledgments}: JWC is supported in part by the MOST, NTU-CTS, and
the NTU-CASTS of R.O.C. J.D. is supported in part by the Major State Basic
Research Development Program in China (Contract No. 2014CB845406), National
Natural Science Foundation of China (Projects No. 11105082).



\begin{thebibliography}{99}
\bibitem{QCDphases} 
M.~A.~Stephanov, 
PoS LAT \textbf{2006}, 024 (2006) [hep-lat/0701002]; 
K.~Fukushima and C.~Sasaki, 
Prog.\ Part.\ Nucl.\ Phys.\ \textbf{72}, 99 (2013) [arXiv:1301.6377
[hep-ph]]; 
S.~Gupta, X.~Luo, B.~Mohanty, H.~G.~Ritter and N.~Xu,
Science {\bf 332}, 1525 (2011)
[arXiv:1105.3934 [hep-ph]]; and references therein.

\bibitem{altlattice} 
P.~de Forcrand and O.~Philipsen, 
Phys.\ Rev.\ Lett.\ \textbf{105}, 152001 (2010) [arXiv:1004.3144 [hep-lat]]; 
A.~Li, A.~Alexandru and K.~F.~Liu, 
Phys.\ Rev.\ D \textbf{84}, 071503 (2011) [arXiv:1103.3045 [hep-ph]].


\bibitem{Gavai:2010zn} R.~V.~Gavai and S.~Gupta, 
Phys.\ Lett.\ B \textbf{696}, 459 (2011) [arXiv:1001.3796 [hep-lat]].


\bibitem{Stephanov:1998dy} M.~A.~Stephanov, K.~Rajagopal and E.~V.~Shuryak, 
Phys.\ Rev.\ Lett.\ \textbf{81}, 4816 (1998) [hep-ph/9806219]. 
M.~A.~Stephanov, K.~Rajagopal and E.~V.~Shuryak, 
Phys.\ Rev.\ D \textbf{60}, 114028 (1999) [hep-ph/9903292].

\bibitem{Aggarwal:2010wy} M.~M.~Aggarwal \textit{et al.} [STAR
Collaboration], 
Phys.\ Rev.\ Lett.\ \textbf{105}, 022302 (2010) [arXiv:1004.4959 [nucl-ex]]. 

\bibitem{Adamczyk:2013dal} L.~Adamczyk \textit{et al.} [STAR Collaboration], 
Phys.\ Rev.\ Lett.\ \textbf{112}, 032302 (2014) [arXiv:1309.5681
[nucl-ex]].

\bibitem{slowing} 
B.~Berdnikov and K.~Rajagopal, 
Phys.\ Rev.\ D \textbf{61}, 105017 (2000) [hep-ph/9912274]. 
C.~Nonaka and M.~Asakawa, 
Phys.\ Rev.\ C \textbf{71}, 044904 (2005) [nucl-th/0410078]. 
C.~Athanasiou, K.~Rajagopal and M.~Stephanov, 
Phys.\ Rev.\ D \textbf{82}, 074008 (2010) [arXiv:1006.4636 [hep-ph]].

\bibitem{Hatta:2003wn} 
  Y.~Hatta and M.~A.~Stephanov,
  Phys.\ Rev.\ Lett.\  {\bf 91}, 102003 (2003)
  [Erratum-ibid.\  {\bf 91}, 129901 (2003)]
  [hep-ph/0302002].


\bibitem{Stephanov:2008qz} M.~A.~Stephanov, 
Phys.\ Rev.\ Lett.\ \textbf{102}, 032301 (2009) [arXiv:0809.3450 [hep-ph]].


\bibitem{Stephanov:2011pb} M.~A.~Stephanov, 
Phys.\ Rev.\ Lett.\ \textbf{107}, 052301 (2011) [arXiv:1104.1627 [hep-ph]].


\bibitem{Asakawa:2009aj} M.~Asakawa, S.~Ejiri and M.~Kitazawa, 
Phys.\ Rev.\ Lett.\ \textbf{103}, 262301 (2009) [arXiv:0904.2089 [nucl-th]].


\bibitem{Skokov:2011rq} V.~Skokov, B.~Friman and K.~Redlich, 
Phys.\ Lett.\ B \textbf{708}, 179 (2012) [arXiv:1108.3231 [hep-ph]].


\bibitem{Schnetz:2005ih} O.~Schnetz, M.~Thies and K.~Urlichs, 
Annals Phys.\ \textbf{321}, 2604 (2006) [hep-th/0511206].

\bibitem{chiMlattice} 
Y.~Aoki, Z.~Fodor, S.~D.~Katz and K.~K.~Szabo, 
Phys.\ Lett.\ B \textbf{643}, 46 (2006) [hep-lat/0609068]. 
A.~Bazavov, T.~Bhattacharya, M.~Cheng, \textit{et al.} 
Phys.\ Rev.\ D \textbf{85}, 054503 (2012) [arXiv:1111.1710 [hep-lat]]. 
L.~Levkova, 
PoS LATTICE \textbf{2011}, 011 (2011) [arXiv:1201.1516 [hep-lat]].
\end{thebibliography}
\end{document}